# Electron-Electron Interactions in 2D Semiconductor InSe

Arvind Shankar Kumar,[1,†] Kasun Premasiri,[1,†] Min Gao,[1] U. Rajesh Kumar,[2] Raman Sankar,[2,3] Fang-Cheng Chou,[2] and Xuan P. A. Gao[1,*]

[1]Department of Physics, Case Western Reserve University, 2076 Adelbert Road, Cleveland, Ohio 44106, USA
[2]Center for Condensed Matter Sciences, National Taiwan University, Taipei 10617, Taiwan
[3]Institute of Physics, Academia Sinica, Taipei 11529, Taiwan
† These authors contributed equally to this work.
*Corresponding author: xuan.gao@case.edu

Electron-electron interactions (EEIs) in 2D van der Waals structures is one of the topics with high current interest in physics. We report the observation of a negative parabolic magnetoresistance (MR) in multilayer 2D semiconductor InSe beyond the low-field weak localization/antilocalization regime, and provide evidence for the EEI origin of this MR behavior. Further, we analyze this negative parabolic MR and other observed quantum transport signatures of EEIs (temperature dependent conductance and Hall coefficient) within the framework of Fermi liquid theory and extract the gate voltage tunable Fermi liquid parameter $F_0^\sigma$ which quantifies the electron spin-exchange interaction strength.

Two-dimensional (2D) van der Waals (vdW) materials offer a versatile platform to venture into new facets of physics, wherein the discovery of exfoliable graphene has been the initial impetus [1–3]. Having a multitude of material candidates with exotic properties that are important in various fields such as topological phases [4–7], spintronics [8–10], valleytronics [11], 2D materials continue to fascinate researchers with unique perspectives. In terms of electron quantum transport, 2D materials provide researchers with a diverse set of materials that can be used to obtain confined 2D electron gases (2DEGs) beyond conventional semiconductor heterostructure systems [12–15]. Quantum transport studies of 2D materials have already shown great promise in various fronts ranging from exploring spin polarization effects and the underlying mechanisms to topological superconductivity and beyond [6–8,16].

Deepening the understanding about electron localization and electron-electron interactions (EEIs) has been one of the major focuses in the studies of disordered 2DEGs [17-36]. Most of these studies are based on conventional semiconductors such as GaAs heterostructures [21-26], Si metal-oxide-semiconductor field-effect transistors (MOSFETs) [27-30] and more recently, graphene [31-33]. The availability and development of new 2D materials in the past few years have opened up the opportunity to study electron quantum transport effects in new materials and/or in different transport regimes. For instance, there have been reports on weak localization (WL), weak antilocalization (WAL), and spin-orbit coupling effects in 2D semiconductors including transition metal dichalcogenides as well as non-transition metal chalcogenides [8,10,37].

However, to the best of our knowledge, a direct observation of EEIs in vdW 2D-semiconductors has not been reported in literature. Probing the EEIs in new 2D materials is compelling due to the diverse set of material choices and electronic band structures to explore, compared to the conventional heterointerfaces based on group IV or III-V semiconductors. To this end, we conduct a comprehensive electron transport study of 2D semiconductor indium monoselenide (InSe) and report here various signatures of EEIs – a negative parabolic MR, and logarithmic temperature (*T*)-dependent Hall coefficient and conductivity [36]. Analyzing these observations within the framework of Fermi liquid (FL) theory in the diffusive transport regime, we are able to extract the interaction parameter $F_0^\sigma$ over the range of electron density $n = 3 - 8 \times 10^{12}/\text{cm}^2$ for InSe, and compare it with the predictions of FL theory.

InSe is a group-III monochalcogenide semiconductor. Se-In-In-Se atoms form the individual vdW layers with a honeycomb lattice structure [15,37,38]. It has drawn a lot of attention due to its relatively high electron mobility among 2D semiconductors [15,38], strong spin-orbit coupling [37] and optoelectronic properties associated with the direct-indirect gap transition as the thickness is reduced [15,39-42]. For this study, InSe nanoflake devices were fabricated on-

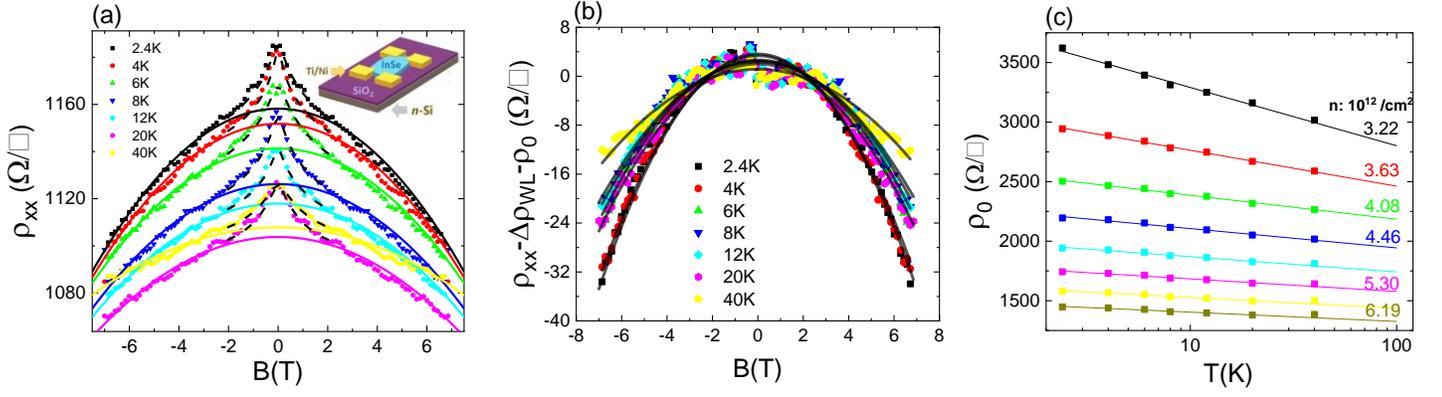

**Fig.1** (a) Magnetoresistance of InSe nanoflake sample 1 at different temperatures, plotted for $n = 6.19 \times 10^{12}/\text{cm}^2$. The dotted lines correspond fittings to the 2D weak localization (WL) effect. The solid lines correspond to fits to parabolas at high fields ($B > 2T$). Inset: device schematic. (b) Magneto-resistance with WL correction subtracted, for data in Fig. 1(a). Black solid lines correspond to parabolic fits. (c) $\rho_0$, which is the $B = 0$ intercept of parabolic fits shown in Fig. 1(a), plotted vs. temperature, at different electron densities.

degenerately-doped Si substrates with 290 nm $SiO_2$ (Sample 1) or 150 nm $Si_3N_4$ (Sample 2) as the gate dielectric. Briefly, β-InSe nanoflakes were mechanically exfoliated onto cleaned substrates and Ti/Ni contacts to the freshly exfoliated flakes were fabricated using stencil lithography with copper grid shadow masks, and electron-beam metal deposition.

The inset of Fig. 1(a) illustrates the device structure and van der Pauw geometry used for transport experiments. Gold wires (50 μm in diameter) were attached to the Ti/Ni metal contacts using Indium soldering, and then mounted onto the sample holder of a Physical Property Measurement System (Quantum Design, Model 6000) and measured with low frequency lock-in techniques. Longitudinal resistivity $\rho_{xx}$ and transverse resistivity $\rho_{xy}$ were calculated using the van der Pauw method [43], followed by symmetrization of the $\rho_{xx}$ vs. magnetic field ($B$) and anti-symmetrization of the $\rho_{xy}$ vs. $B$ data to remove mixings between $\rho_{xx}$ and $\rho_{xy}$ due to imperfect contact alignment.

Fig 1 and Fig 2 show the magnetoresistance (MR) and Hall effect data for our InSe Sample 1 (all data presented in the main text are from Sample 1 unless stated). In Fig.1(a), we present the longitudinal MR data which exhibit the WL behavior at low fields ($B < B_l = \hbar/4eD\tau \sim 2$ T for our system with $D$ as the diffusion constant) [21,44] and changes into a negative parabolic-like MR at higher fields. WL or WAL effect [45–47] are known to be induced by the single particle quantum interference corrections for electrons diffusing in a random impurity potential and leads to a negative (for WL) or positive (for WAL) MR [44,48,49] in the weak magnetic field limit ($\omega_C \tau \ll 1$), where $\omega_C$ is the cyclotron frequency ($\omega_C = eB/m^*$) and $\tau$ is the momentum relaxation time. However, the cross-over to a different type of negative MR above the critical field $B_l$ as observed here is unusual compared to similar analyses performed in other 2D vdW materials [8,16].

To further explore this negative parabolic MR behavior at $B > B_l$, we subtract the WL correction $\Delta\rho_{WL}(B)$ which predominates at $B < B_l$. $\Delta\rho_{WL}(B)$ is determined by fitting the low field data ($B < 2T$) to the HLN equation for 2D WL [48]. $\rho_{xx}(B) - \Delta\rho_{WL}(B) - \rho_0$ is plotted in Fig. 1(b) with $\rho_0$ being the zero-field resistivity without the WL effect. A negative parabolic dependence is clearly observed.

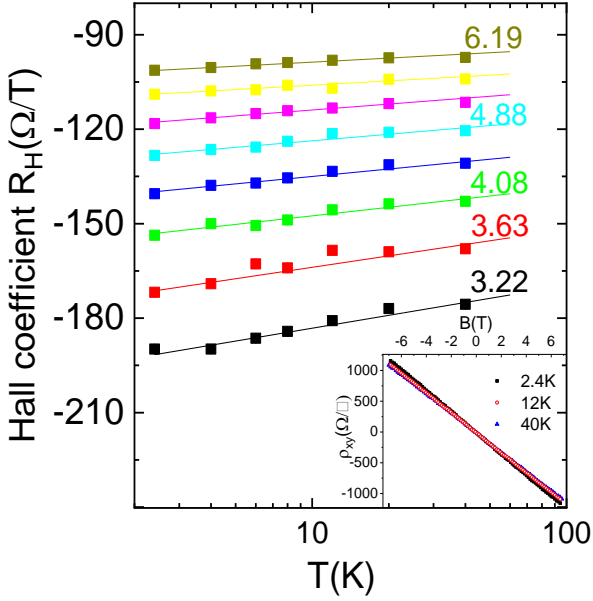

**Fig.2** Temperature dependence of Hall coefficient $R_H$ at different electron densities (values marked on curves in units $10^{12}$/cm$^2$). Inset: Example transverse/Hall resistivity $\rho_{xy}$ vs. $B$ data at different temperatures, plotted for $n = 3.63 \times 10^{12}$ /cm$^2$.

$\rho_0$ is obtained as the $B = 0$ intercept in fitting the high field ($B > B_l$) data in Fig.1(a) to a parabolic dependence. The temperature dependence of $\rho_0$ is shown in Fig.1(c). Strikingly, it exhibits a logarithmic $T$-dependence, even with the WL effect removed. This is a consequence of EEI effect as discussed later.

Fig.2 shows the $T$-dependence of Hall coefficient, $R_H$, extracted from linear fit to $\rho_{xy}$ vs. $B$ (shown in inset). We observe a similar logarithmic behavior to this data as well. These data are clear signatures of the effect of EEIs for a 2D FL in our InSe system. In conventional semiconductor heterointerfaces, EEIs were found to give rise to a negative parabolic MR effect at magnetic fields where Zeeman splitting effect is negligible [25,36], along with logarithmic corrections to $T$-dependence of zero field conductance and Hall coefficient [33,36]. While most prior transport studies on conventional semiconductor heterostructures and graphene are in the ballistic ($\frac{k_B T \tau}{\hbar} > 1$) and high field ($\omega_c \tau > 1$) regime, the gated InSe nanoflake here resides in the diffusive ($\frac{k_B T \tau}{\hbar} < 1$) and low-field ($\omega_c \tau < 1$) regime. In the next section, we outline three methods of extracting $\delta\sigma_{xx}^{ee}(T)$, the EEI correction to the Drude conductivity from these transport data.

The negative parabolic MR due to EEIs is given by

$$\rho_{xx} = \frac{1}{\sigma_0}\left[1 - \frac{(1-(\omega_c\tau)^2)\delta\sigma_{xx}^{ee}}{\sigma_0}\right] \quad (1)$$

Where $\omega_c$ is the cyclotron frequency and $\sigma_0$ is Drude conductivity [36]. The correction to $R_H$ is given by

$$\frac{\Delta R_H}{R_H^0} = -\frac{2\Delta\sigma_{xx}^{ee}}{\sigma_0} \quad (2)$$

Where $R_H^0$ is the classical Hall coefficient [33,36]. These relations are derived by inverting the 2D conductivity tensor $\boldsymbol{\sigma}$ with corresponding EEI corrections added (Supplementary Info). For FL theory in the diffusive limit ($\tau \ll \frac{\hbar}{k_B T}$), $\delta\sigma_{xx}^{ee}(T)$ is known to have a logarithmic $T$-dependence given by

$$\delta\sigma_{xx}^{ee} = \frac{e^2}{2\pi^2\hbar} f(F_0^\sigma) \ln\left(\frac{T}{T_0}\right) \quad (3)$$

where $T < T_0 = \frac{\hbar}{k_B\tau}$ and $f(F_0^\sigma) = 1 + 3(1 - \ln(1 + F_0^\sigma)/F_0^\sigma)$ [50,51].

We can extract $\delta\sigma_{xx}^{ee}(T)$ from fitting the data in Fig.1(b) to the $B^2$ dependence according to Eq.1 (Method 1), or comparing $\rho_0(T)$ shown in Fig.1(c) to the $B = 0$ limit of Eq. 1 (Method 2), or fitting the Hall coefficient data shown in Fig.2 to Eq.(2) (Method 3). The results of these methods are shown in Fig.3. From the extracted $\delta\sigma_{xx}^{ee}(T)$, we observe a logarithmic divergence as $T$ is lowered in all three methods, as expected from Eq.(3). This provides further evidence for the EEI source of these corrections. We can then use the logarithmic dependence of $\delta\sigma_{xx}^{ee}(T)$ to extract the FL parameter $F_0^\sigma$ from Eq.(3). The results are plotted in Fig.4. It is important to note that, even though the order of magnitude of the corrections and the $T$-dependence are consistent across all three methods, we observe an approximate factor of three higher magnitude in the $\delta\sigma_{xx}^{ee}(T)$ extracted from MR (Methods 1 and 2) compared to the Hall coefficient analysis (Method 3).

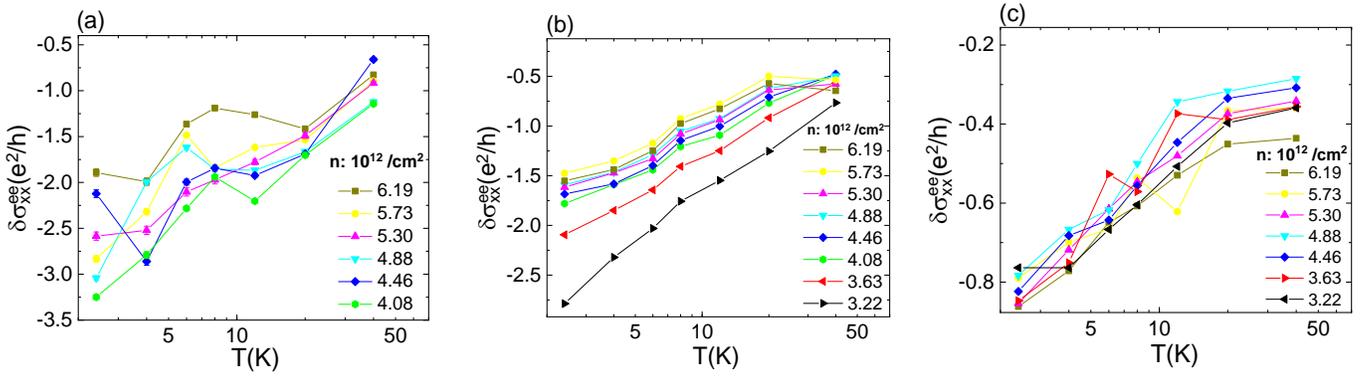

**Fig.3** EEI correction to Drude conductivity vs. $T$, extracted from (a) negative parabolic MR (Method 1); (b) zero field resistance without the WL effect, given by $\rho_0$ (Method 2); (c) change in Hall coefficient with temperature (Method 3).

This enhancement is also reflected in their larger slope of logarithmic $\delta\sigma_{xx}^{ee}(T)$ (Supplementary Info), which is the relevant quantity for determining the FL parameter $F_0^\sigma$. Even though the difference in the slope of $\delta\sigma_{xx}^{ee}(T)$ is mostly within a factor of 3 across all methods, this difference becomes significant when extracting $F_0^\sigma$, due to the highly non-linear dependence of the logarithmic slope of $\delta\sigma_{xx}^{ee}(T)$ on $F_0^\sigma$ (Supplementary Info). Next, we discuss the extracted $F_0^\sigma$ from all three methods and compare to FL theory. We observe that $F_0^\sigma$ extracted from the Hall effect using Method 3 (Fig.4) shows agreement with predictions of FL theory [50]:

$$F_0^\sigma = -\frac{1}{2\pi}\left(\frac{r_s}{\sqrt{2-r_s^2}}\right)\ln\left[\frac{\sqrt{2}+\sqrt{2-r_s^2}}{\sqrt{2}-\sqrt{2-r_s^2}}\right] \quad (4)$$

where $r_s = 1/\sqrt{\pi(a_B^*)^2 n}$ is the 2D interaction parameter with $a_B$ as the effective Bohr radius and $n$ being the electron density. This prediction is valid for $r_s^2 < 2$, and we find that our system satisfies this inequality in the density range considered.

While $F_0^\sigma$ extracted from Hall effect agrees with FL theory, $F_0^\sigma$ extracted from MR data (Methods 1 and 2) show strong density dependences not consistent with FL theory (inset of Fig.4). A possible reason for the observed discrepancy in $F_0^\sigma$ extracted from MR, which is reflected in the enhanced $\delta\sigma_{xx}^{ee}(T)$ extracted from MR, may be the non-uniform current flow in the van der Pauw geometry, leading to errors in determining $\rho_{xx}$ of our sample. Even though we averaged our measurements across different contact configurations, and accounted for observed anisotropies in resistance between different contact configurations (according to [43]), errors due to finite size of contacts and non-ideal sample shape would still possibly lead to further deviations from the ideal van der Pauw resistance pre-factor of $\pi/Ln(2)$ [43].

It is also possible that other known effects contribute to negative MR but it is unlikely that these effects are relevant to InSe in this transport regime. In electronic systems with high carrier mobility, the parabolic negative MR is complemented by a classical contribution, at strong magnetic fields. The presence of an external magnetic field allows electrons to become localized around impurities [52,53]. However, the relevance of this effect in here can be neglected considering the carrier mobility values (~1000 cm$^2$/Vs at 10K) of the system. The Kondo effect which explains the scattering of conduction electrons due to magnetic impurities can also lead to negative MR [54,55], but this effect also becomes irrelevant in this context since InSe has been predicted to be non-magnetic [56]. The chiral anomaly in Dirac/Weyl semimetals also cause negative MR, where the application of $E||B$ breaks the chiral symmetry, and as a consequence, the conservation of chiral charge density is violated by a quantity called the anomaly term [57,58]. But this effect is inapplicable to InSe, which is a semiconductor.

In summary, we observe a negative parabolic MR effect in exfoliated InSe nanoflakes. We attribute this MR to EEI effects, as supported by the concomitant observation a logarithmic $T$-dependent conductivity and Hall coefficient. We extract the EEI correction to Drude conductivity and the FL parameter $F_0^\sigma$ within the framework of FL theory in the diffusive transport regime. We find that $F_0^\sigma$ extracted from

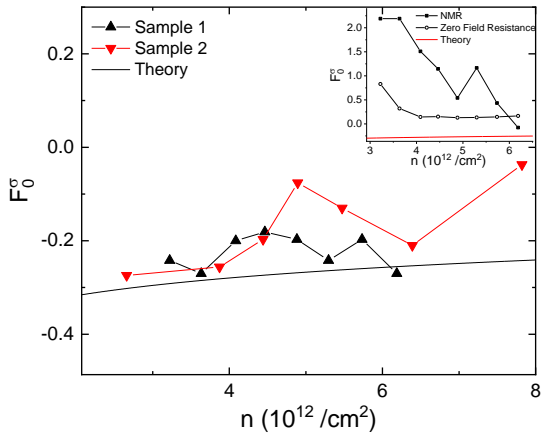

**Fig.4** FL parameter extracted using Hall coefficient (Method 3) plotted versus electron density. Solid line corresponds to prediction of FL theory (Eq.4). Inset: FL parameter extracted using Method 1 and Method 2 (for Sample 1) versus electron density.

Hall effect analysis shows good agreement over the density range $(3 - 7 \times 10^{12}/\text{cm}^2)$ to FL theory. To the best of our knowledge, this is the first transport study of EEI effects in 2D vdW semiconductors. Our study shows the possibility of using InSe and other 2D vdW semiconductors and heterostructures as a platform for studying 2D electron physics in general.

XPAG acknowledges the financial support from NSF (DMR-1607631). FCC and RS acknowledge funding support from the Ministry of Science and Technology (108-2622-8-002-016 and 108-2112-M-001-049-MY2), from the Ministry of Education in Taiwan (AI-MAT 108L900903), and from the Academia Sinica (AS-iMATE-108-11).

# Supplementary Information for 'Electron-Electron Interactions in 2D Semiconductor InSe'

Arvind Shankar Kumar,[1,†] Kasun Premasiri,[1,†] Min Gao,[1] U. Rajesh Kumar,[2] Raman Sankar,[2,3] Fang-Cheng Chou,[2] and Xuan P. A. Gao[1,*]

[1]Department of Physics, Case Western Reserve University, 2076 Adelbert Road, Cleveland, Ohio 44106, USA
[2]Center for Condensed Matter Sciences, National Taiwan University, Taipei 10617, Taiwan
[3]Institute of Physics, Academia Sinica, Taipei 11529, Taiwan
† These authors contributed equally to this work.
*Corresponding author: xuan.gao@case.edu

## 1. Device details

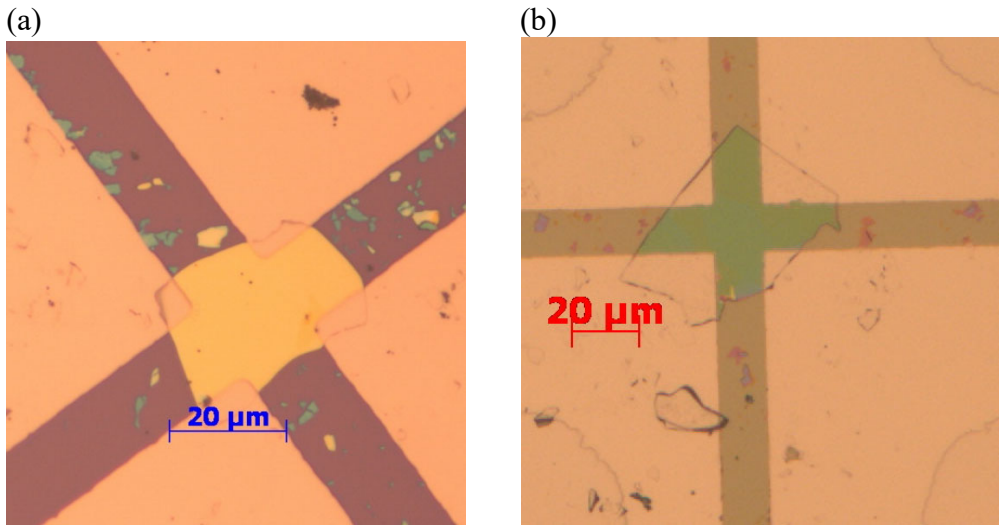

| Sample | Substrate | Thickness of InSe flake |
|---|---|---|
| Sample 1 | 300 nm $SiO_2$ /$n$-Si | 55 nm |
| Sample 2 | 150 nm $Si_3N_4$ /$n$-Si | 75 nm |

**Fig.S1** Optical image of (a) Sample 1 and (b) Sample 2. Table shows information on substrates and corresponding thickness of InSe flakes used, measured by AFM.

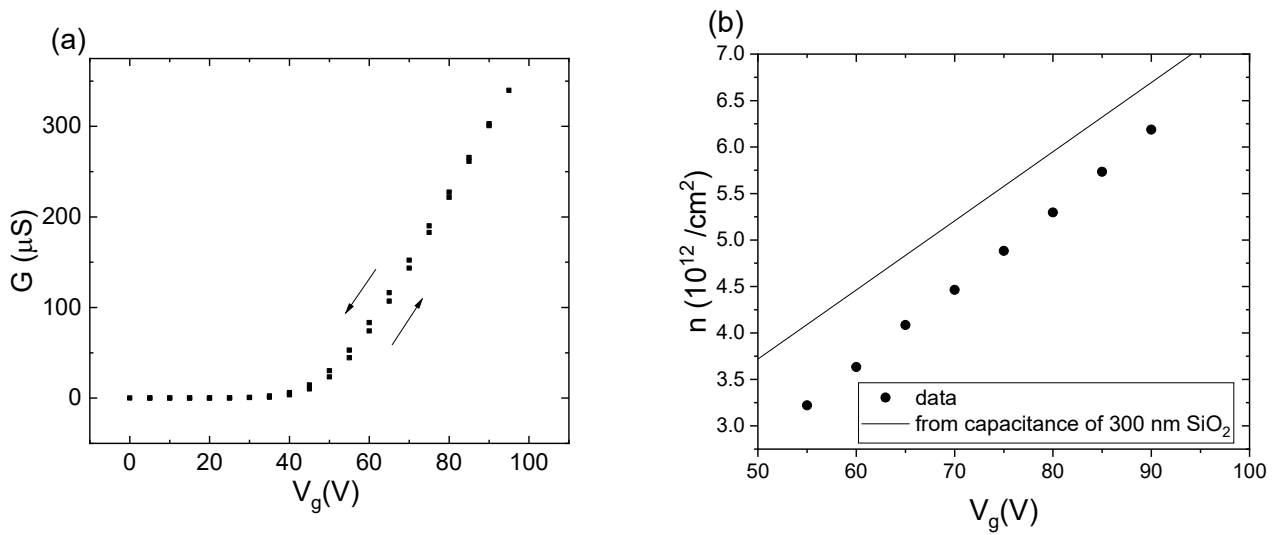

**Fig.S2** Transport characteristics of Sample 1 at T=10K. S2(a) shows variation of two-probe conductance over a range of gate voltages, measured using low-frequency lock-in technique with a source-drain voltage of 4 mV. S2(b) shows variation of carrier density over the range of gate voltages studied in main paper, extracted from Hall effect data.

## 2. Sample 2 Magnetoresistance and Hall effect data

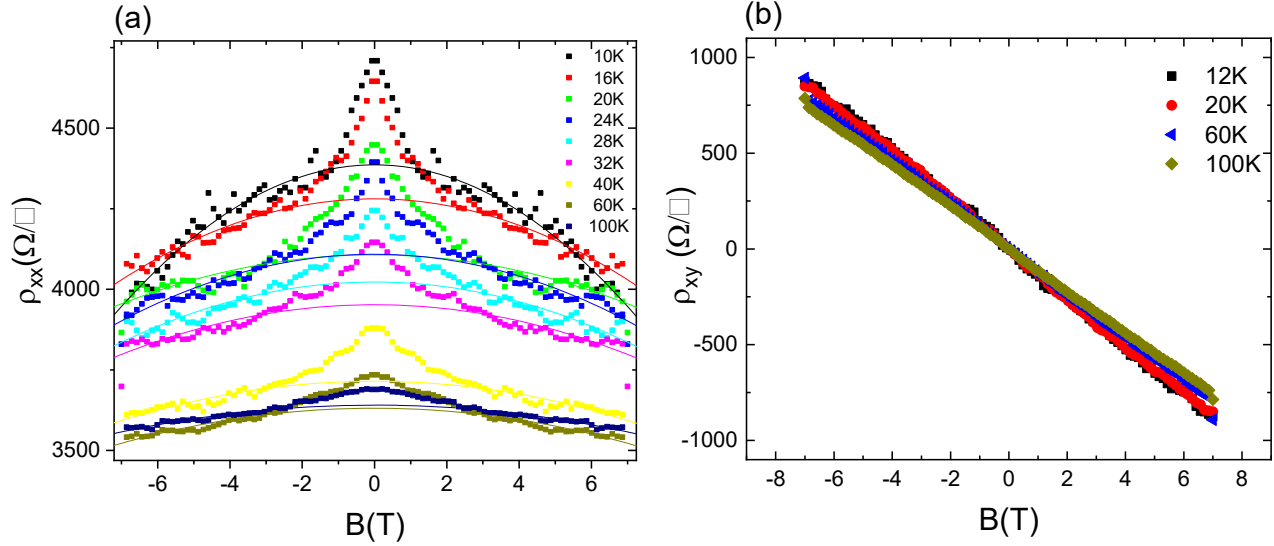

**Fig.S3** (a) Magnetoresistance data for Sample 2 at different temperatures for $n = 4.7 \times 10^{12}\ /cm^2$. Solid lines correspond to fits to $B^2$ dependence. (b) Hall effect data for Sample 2 at different temperatures for the same electron density.

## 3. EEI correction to Drude conductivity

Consider the conductivity tensor

$$\boldsymbol{\sigma} = \begin{bmatrix} \sigma_{xx} & \sigma_{xy} \\ -\sigma_{xy} & \sigma_{xx} \end{bmatrix} \qquad (1)$$

The elements are given by

$$\sigma_{xx} = \frac{\sigma_0}{(1+(\mu B)^2)} + \delta\sigma_{xx}^{ee} \qquad (2)$$

and

$$\sigma_{xy} = \frac{\sigma_0(\mu B)}{(1+(\mu B)^2)} \qquad (3)$$

We get the resistivity tensor $\boldsymbol{\rho} = \begin{bmatrix} \rho_{xx} & \rho_{xy} \\ -\rho_{xy} & \rho_{xx} \end{bmatrix}$ by inverting $\sigma$. Its elements are given by

$$\rho_{xx} = \frac{1}{\sigma_0}\left[1 - \frac{(1-(\mu B)^2)\delta\sigma_{xx}^{ee}}{\sigma_0}\right] \qquad (4)$$

$$\rho_{xy} = -\frac{(\mu B)}{\sigma_0}\left[1 - \frac{2\delta\sigma}{\sigma_0}\right] \qquad (5)$$

Since $\mu B = \omega_c \tau$, Eq.(4) and Eq.(5) are the same as Eq.(1) and Eq.(2) in main text.

## 4. EEI interaction parameter

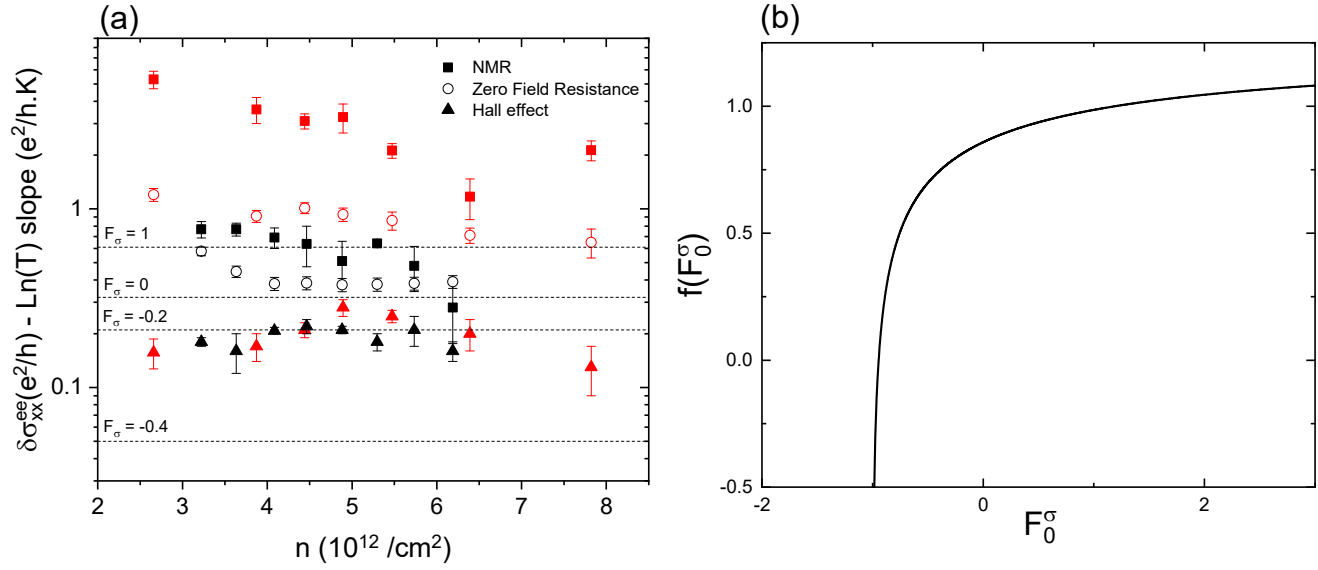

Fig.S4 (a) Logarithmic slope of $\delta\sigma_{xx}^{ee}(T)$ plotted against electron density for Sample 1 (black) and Sample 2 (red). Data is plotted for $\delta\sigma_{xx}^{ee}(T)$ extracted from all three methods described in main text. (b) Relationship between $f(F_0^\sigma) = \frac{1}{\pi}(1 + 3(1 - \ln((1 + F_0^\sigma)/F_0^\sigma)))$ and $F_0^\sigma$ is shown over a range of $F_0^\sigma$ values.

Fig.S4(a) shows the logarithmic slope of $\delta\sigma_{xx}^{ee}(T)$ plotted against electron density for Sample 1 (black) and Sample 2 (red). The interaction parameter $F_0^\sigma$ depends directly on this value through Eq.(3) in main text. For Sample 2, the MR correction from $\delta\sigma^{WL}(B)$ has not been considered due to low-field data being too noisy for reliable fitting to the HLN equation. This is likely the reason for the slopes extracted from MR and zero-field resistance for Sample 2 (red squares and circles) being higher than those of Sample 1. Data extracted from Hall effect are consistent for both samples and have been used in the main text.